\PassOptionsToPackage{hyphens}{url}

\documentclass[a4paper, 10pt, conference]{ieeeconf}      

\IEEEoverridecommandlockouts                              

\usepackage{comment}
\usepackage{hyperref}
\usepackage{graphicx}
\usepackage{color}
\usepackage{amsmath}
\usepackage{amssymb}
\usepackage{xcolor}
\usepackage{color}
\usepackage{float}
\usepackage{subcaption}
\usepackage{listings}
\usepackage{todonotes}
\usepackage{xurl}
\usepackage{fltrace}
\usepackage{lipsum}

\colorlet{punct}{red!60!black}
\definecolor{background}{HTML}{EEEEEE}
\definecolor{delim}{RGB}{20,105,176}
\colorlet{numb}{magenta!60!black}

\lstdefinelanguage{json}{
    basicstyle=\normalfont\ttfamily,
    numbers=left,
    numberstyle=\scriptsize,
    stepnumber=1,
    numbersep=8pt,
    showstringspaces=false,
    breaklines=true,
    frame=lines,
    backgroundcolor=\color{background},
    literate=
     *{0}{{{\color{numb}0}}}{1}
      {1}{{{\color{numb}1}}}{1}
      {2}{{{\color{numb}2}}}{1}
      {3}{{{\color{numb}3}}}{1}
      {4}{{{\color{numb}4}}}{1}
      {5}{{{\color{numb}5}}}{1}
      {6}{{{\color{numb}6}}}{1}
      {7}{{{\color{numb}7}}}{1}
      {8}{{{\color{numb}8}}}{1}
      {9}{{{\color{numb}9}}}{1}
      {:}{{{\color{punct}{:}}}}{1}
      {,}{{{\color{punct}{,}}}}{1}
      {\{}{{{\color{delim}{\{}}}}{1}
      {\}}{{{\color{delim}{\}}}}}{1}
      {[}{{{\color{delim}{[}}}}{1}
      {]}{{{\color{delim}{]}}}}{1},
}

\title{\LARGE \bf Cyber-Physical Digital Factory Architecture\\ as the Enabler of Disembodied Work}

\author{Tero Kaarlela$^{1}$, Ivan Ruchkin$^{2}$, Jose Outeiro$^{3}$ and Souradeep Dutta$^{4}$
\thanks{$^{1}$Materials and Mechanical Engineering, Faculty of Technology, University of Oulu, Oulu, Finland.
        {\tt\small tero.kaarlela@oulu.fi}}%
\thanks{$^{2}$Electrical and Computer Engineering, University of Florida, Gainesville, FL, USA.
       }%
\thanks{$^{3}$Digital Engineering for Advanced Manufacturing Laboratory (DEAM Lab), Center for Precision Metrology, Department of Mechanical Engineering and Engineering Science, University of North Carolina at Charlotte, 9201 University City Blvd., Charlotte 28223, NC, USA
        }%
\thanks{$^{4}$
        Electrical and Computer Engineering, University of British Columbia, Vancouver, BC, Canada
        }%
\thanks{This work has been submitted to the IEEE for possible publication.
Copyright may be transferred without notice, after which this version may
no longer be accessible.}
}

\begin{document}

\tracefloats

\maketitle
\thispagestyle{empty}
\pagestyle{empty}

\begin{abstract}

Digital Twins (DTs), Artificial Intelligence (AI), and Industrial Internet of Things (IIoT) technologies have significantly advanced manufacturing digitalization. However, these technologies are typically applied to individual manufacturing processes rather than integrated into a unified cyber-physical manufacturing environment. This paper proposes a cyber-physical digital factory architecture that enables disembodied work, where manufacturing systems can be supervised and operated remotely through eXtended Reality (XR) user interfaces in collaboration between AI-based control and human operators. The architecture integrates synchronized DTs, hierarchical cloud-edge AI, IIoT, and XR teleoperation interfaces into a cyber-physical manufacturing environment. The proposed approach is validated through representative manufacturing operations, including CNC machining, robotic-assisted abrasive finishing, and robotized disassembly. The results demonstrate the feasibility of the proposed architecture for disembodied manufacturing work and provide a reusable cyber-physical framework for future human-AI-controlled digital factories.

\end{abstract}

\section{INTRODUCTION}
\label{sec:introduction}

\subsection{Motivation}
\label{subsec:motivation}
The manufacturing and construction sector faces challenges in workforce availability and offshoring of production to lower-cost countries~\cite{Denisa2023technology, robert2020growing}. Younger generations perceive factory work as inflexible, unsafe, and non-ergonomic~\cite{Mohan2024genz, Paradis2024genz}, with surveys showing that 77\% would consider leaving a job requiring full on-site presence~\cite{Stephenson2025digital} and 21\% would accept a pay cut~\cite{bartik2023rise} to maintain remote work. While digital office environments expanded during the pandemic, remote work is rarely possible for production machine operators, creating an imbalance among workers and further reducing the attractiveness of manufacturing jobs~\cite{wullf2024rise}. 

Trade exposure illustrates the fragility of manufacturing employment: between 2001 and 2018, the U.S.--China trade deficit is associated with the loss of approximately 3.7 million U.S. jobs, nearly 75\% of which were in manufacturing~\cite{robert2020growing, Moffat2021big}. To ensure workforce availability and profitability, manufacturing must enable location-independent work to improve flexibility, safety, and ergonomics. Competitiveness in high-cost countries will increasingly rely on agile, small-lot production, where enhanced flexibility is essential for both profitability and sustainability~\cite{Denisa2023technology,Jin2023achieving}.

Despite significant advances in manufacturing automation, skilled operators remain physically tied to manufacturing facilities because supervision, exception handling, and complex decision-making still require direct interaction with manufacturing resources~\cite{Kim2024teleoperator, Kaarlela2025evb}. As a result, manufacturing cannot benefit from the same location-independent work practices that have transformed knowledge-intensive professions. This limits workforce flexibility, restricts access to geographically distributed expertise, and reduces the attractiveness of manufacturing jobs.

Addressing this challenge requires a manufacturing paradigm that enables safe, flexible, and \textit{location-independent} operation of manufacturing resources. Such a paradigm extends manufacturing towards Cyber-Physical Systems (CPSs) in which physical manufacturing resources, DTs, AI agents, and human operators interact through synchronized feedback loops. This paper describes a technological vision for \textit{disembodied work}, which refers to manufacturing activities in which humans supervise, collaborate with, or intervene in physical manufacturing systems without being co-located with the manufacturing resources. Unlike conventional teleoperation, disembodied work allows dynamic allocation of decision authority between autonomous systems and remote human operators through synchronized DTs. While \textit{disembodied work} does not itself prevent offshoring, it keeps physical capital, tooling, material flows, and the resulting production onshore, while widening the labor pool that can operate it.

\subsection{State-of-the-art}
\label{sec:relatedresearch}
Early concepts of the digital factory envisioned a unified digital environment in which engineering tools, production planning, simulation models, and factory operation were integrated through a common information infrastructure~\cite{Bracht2005the,Bracht2018digitale}. The digital factory was expected to evolve together with its physical counterpart, enabling simulation, monitoring, optimization, and lifecycle management of manufacturing systems~\cite{Debevec2022digital}. Recent developments have further expanded this vision toward cyber-physical manufacturing environments integrating DTs, the IIoT, cloud-edge computing, AI, and real-time data analytics~\cite{Burggraf2024towards}.

DTs have become key enablers of digital factories by providing continuous synchronization between physical and digital machines, processes, and manufacturing systems. Recent research has demonstrated DTs for machining process monitoring, hybrid additive-subtractive manufacturing, predictive maintenance, quality assurance, and optimization~\cite{Tao2024advancements}. These capabilities enable manufacturing resources to be continuously monitored, analyzed, and controlled.

AI has further expanded digital manufacturing by enabling intelligent process monitoring, anomaly detection, predictive maintenance, production scheduling, autonomous operation, and decision-making support. Together with IIoT real-time data, AI enables manufacturing systems to react dynamically to changing conditions, increasing the level of autonomous manufacturing operations~\cite{Eichelberger2025industry}.

Teleoperation complements autonomous manufacturing by enabling location-independent supervision and control through remote human interaction~\cite{Kim2024teleoperator}. Recent advances in eXtended Reality (XR) and human--robot interaction have improved operator situational awareness and remote manipulation capabilities, making teleoperation an increasingly viable solution for manufacturing applications~\cite{Wan2024virtual}.

From a Cyber-Physical System (CPS) perspective, these technologies represent complementary capabilities required to couple cyber and physical domains. DTs provide CPS state representation, IIoT enables information exchange, AI supports decision-making, and teleoperation enables human participation. Together, they enable manufacturing systems in which physical processes and systems, and computational intelligence operate as a unified CPS.

\subsection{Reference Architectures and Standards} 
\label{subsec:standards} 
Several reference architectures address parts of the design space considered in this work. ISO 23247 defines a manufacturing DT framework~\cite{ISO23247part2}. The 5C architecture~\cite{Lee2015cyber} provides a CPS reference model; RAMI 4.0 and the Asset Administration Shell~\cite{IEC63278} support interoperability; IEC 62264~\cite{IEC62264} defines production-management functions; and recent DT orchestration platforms support cloud-edge composition of heterogeneous services. Table~\ref{tab:comparison} positions the proposed architecture against these approaches. 

\begin{table}[h!tb] 
\centering 
\caption{Positioning of the proposed architecture. \checkmark = primary focus, $\sim$ = partially addressed, -- = not fully addressed.} 
\label{tab:comparison} 
\scriptsize 
\begin{tabular}{lcccc} \hline 
\textbf{Framework} & \textbf{DT} & \textbf{Orch.} & \textbf{H--AI} & \textbf{Safety} \\ 
\hline ISO 23247 & \checkmark & \checkmark & -- & -- \\ 
5C Architecture & \checkmark & $\sim$ & -- & -- \\ 
RAMI 4.0 / AAS & \checkmark & \checkmark & -- & $\sim$ \\ 
IEC 62264 & -- & \checkmark & -- & -- \\ 
DTCL & \checkmark & \checkmark & -- & -- \\ 
Digital Factory & $\sim$ & \checkmark & -- & -- \\ \textbf{This work} & \checkmark & $\sim$ & \checkmark & \checkmark \\ \hline 
\end{tabular} 
\end{table}

\subsection{Research Gap}
\label{subsec:researchgap}
The mentioned technologies enabling digital factories have primarily evolved as independent research domains or have been applied to individual manufacturing processes, cells, or systems. Existing digital factory implementations mainly support production planning and simulation, while DT research focuses on manufacturing processes and systems. Our previous work contributed toward location-independent manufacturing by developing CPSs supporting DTs for remote monitoring and teleoperation of individual manufacturing processes and systems~\cite{Kaarlela2025evb,Kaarlela2025cyber, outeiro_development_2025}. 

However, a unified operational CPS architecture that integrates heterogeneous manufacturing resources at the factory scale, manufacturing process/system DTs, AI-driven monitoring and control, and human supervision and teleoperation into a unified digital factory has not been established. Consequently, disembodied work through seamless collaboration between autonomous manufacturing systems and remote operators remains largely unsupported.


The scientific contributions of this work are: (1) A cyber-physical digital factory architecture that extends traditional simulation-oriented digital factories by integrating DTs, AI, autonomous manufacturing systems, and human supervision and teleoperation into a continuously synchronized cyber-physical manufacturing environment. (2) A human-in-the-loop execution framework enabling dynamic allocation of control authority between autonomous manufacturing systems and remote operators. (3) Validation of the architecture capabilities through CNC machining, robotic-assisted finishing, and Electric Vehicle Battery (EVB) disassembly.

\section{PROPOSED DIGITAL FACTORY CONCEPT}
\label{sec:proposedconcept}
We aim to extend the traditional concept of simulation-oriented digital factories toward an operational cyber-physical manufacturing environment to support disembodied work. Unlike conventional manufacturing systems, where operators and manufacturing resources are co-located, disembodied work requires geographically distributed humans, intelligent software, and physical manufacturing systems to function as a single coordinated manufacturing environment. Achieving this requires addressing several technical challenges related to interoperability, autonomous operation, communication, and cybersecurity.

Coordinating multiple manufacturing resources across manufacturing workflows requires continuously synchronized DTs and a common communication framework capable of supporting real-time information exchange. Furthermore, autonomous manufacturing systems must collaborate seamlessly with remote human operators, allowing control to transition safely between AI and human teleoperation whenever task uncertainty, perception accuracy, or safety constraints exceed predefined limits. Since manufacturing resources may be connected through geographically distributed cloud services, the architecture must also tolerate communication latency and network disruptions while maintaining safe operation. Finally, secure communication, authenticated access, and protection against cyberattacks are essential prerequisites for operating remotely accessible manufacturing systems.

The proposed architecture addresses these challenges through the integration of synchronized DTs, hierarchical cloud-edge AI, resilient communication, and human supervision and teleoperation into a unified cyber-physical manufacturing environment. 

\subsection{Digital Factory Components} 
\label{subsec:digitalfactorycomponents}
\begin{figure}[h!tb]
  \centering
  \includegraphics[width=\columnwidth]{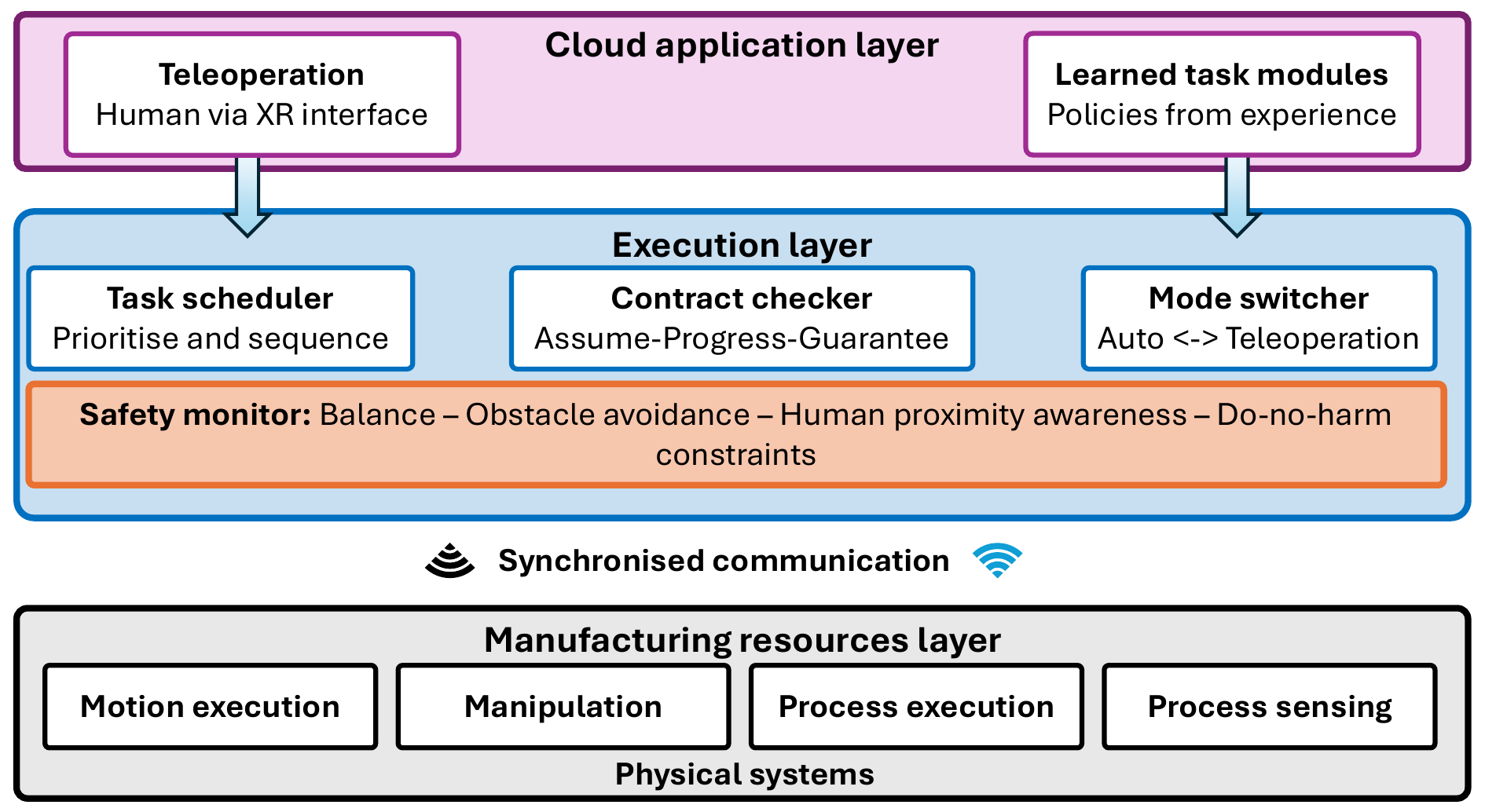}
  \caption{Proposed three-layer architecture for disembodied manufacturing work.}
  \label{fig:PiOS}
\end{figure}

The proposed architecture organizes cyber-physical functionality into three layers illustrated in Figure~\ref{fig:PiOS} operating at different temporal and decision-making scales. Physical manufacturing resources execute low-level actions (manufacturing resources layer), edge intelligence provides local perception and control (execution layer), cloud services coordinate factory-level optimization, and human operators provide supervisory decision-making through teleoperation (cloud application layer). Together, these layers form interconnected cyber-physical feedback loops spanning machine, manufacturing cell, and factory levels.

The \textbf{cloud application layer} is where the human interacts with the system using the XR user interfaces. Interaction supports two modes of operation. In teleoperation mode, an operator controls the equipment in real time through an XR user interface. This is used for novel or unstructured tasks that have not been encountered before. For instance, disassembling an unknown type of EVB or handling a CNC machine error condition. In autonomous mode, the system draws on learned task modules and Reinforcement Learning (RL) based policies built from prior demonstrations and operational data to execute familiar tasks without continuous human supervision. For instance, repeated disassembly of known-type EVBs or machining parts under normal operating conditions. These two modes are not mutually exclusive; The operator can hand off to automation mid-task and reclaim teleoperation at any point. 

The \textbf{execution layer} is the operational core, which has three components working in concert. The task scheduler coordinates concurrent cyber-physical activities across various manufacturing resources. To ensure safe operation, every activity is represented as a contract consisting of assumptions, progress conditions, and guarantees. The contract checker continuously verifies these conditions during execution, enabling runtime validation of cyber-physical state transitions and preventing conflicting operations. An action may only begin if its preconditions are met, must show measurable progress while active, and must satisfy its postconditions before the system moves on. 

The \textbf{mode switcher} governs dynamic allocation of decision authority between autonomous control and human supervision and teleoperation. This enables a human-in-the-loop CPS architecture in which control responsibility can migrate between AI agents (autonomous), hybrid mode (semi-autonomous), and full remote operation according to task complexity, confidence levels, safety constraints, and communication conditions. A practical realization of this arbitration is well established in the CPS community. The formal analysis of this setting is well established and is known as the  Simplex architecture \cite{simplex_lui_sha}. More generally, runtime assurance involves a decision module handing control from a high-performance but unverified controller to a verified safe controller whenever the system approaches the boundary of the region from which safety can still be recovered. Our \textbf{mode switcher} adopts the same structure. It treats the autonomous policy as the high-performance controller and human-in-the-loop control as the safe fallback. Since a human takes over instead of an automatic controller, the safety guarantee also depends on how quickly the operator can respond.

Running across all three, at all times, is the \textbf{safety monitor}. It is not tied to any specific task. It enforces collision avoidance, safe operational linear speed, and protection distance to human workers and other machines continuously and independently of other concurrently running tasks.

The \textbf{Manufacturing Resources Layer} comprises the physical manufacturing systems and exposes standardized capabilities for motion execution, manipulation, process execution, and process sensing. By providing a common abstraction of heterogeneous manufacturing resources, the layer enables the upper execution and cloud application layer to coordinate diverse manufacturing systems through a unified operational interface.

\subsection{Cyber-Physical State Synchronization} 
\label{subsec:dataflowsynchronization}
Continuous synchronization of cyber and physical states is a prerequisite for disembodied work. The proposed low-level communication provides continuous synchronization between physical manufacturing resources, DTs, cloud services, and remote operators. To decouple software components and enable scalable integration of heterogeneous manufacturing systems, the proposed architecture adopts a publish/subscribe communication model implemented using the Message Queuing Telemetry Transport (MQTT) protocol.  

\subsection{Artificial Intelligence Operation} 
\label{subsec:artificialintelligencesupportedoperation}
The proposed digital factory employs a hierarchical AI architecture consisting of cloud-based and edge-based intelligence. Within the proposed CPS architecture, AI functions as a distributed decision-making support layer. Cloud AI operates at the orchestration level, while edge AI closes low-latency perception-action loops near physical manufacturing systems. 

\textbf{Cloud AI} performs factory-level functions, including resource allocation, scheduling, operator management, long-term knowledge management, and factory-wide decision-making. This layer works at the level of higher-level work modules, which capture the description of the manufacturing process, rather than how those tasks are accomplished. 

\textbf{Physical AI} is deployed alongside manufacturing systems/processes to provide the low-latency perception, reasoning, and execution required for autonomous operation. In the robotized EVB disassembly process, it combines machine vision and machine learning to detect and localize components, enabling an industrial robot to autonomously position and actuate the gripper for component detachment~\cite{Karami2025autonomous}. The building blocks for this are well established in the robot learning literature, primarily through the training of large transformer-based models such as Vision-Language-Action (VLA) models \cite{sridhar2025regent}. The architecture has been shown to be effective at learning sequential decision-making tasks from human demonstrations, and then generalizing beyond the training set.

\subsection{Human-in-the-Loop operation}
\label{subsec:humanintheloop}
Humans remain an integral part of the proposed digital factory. Autonomous manufacturing continues while predefined confidence and safety criteria are satisfied. Whenever these criteria are violated, control is seamlessly transferred to a remote operator through the XR teleoperation interface.

Human supervision and teleoperation are required when the AI has reached its accuracy or decision-making limits. In the EVB disassembly process, if detaching or identifying components fails due to corroded or broken screws, autonomous operation is paused until a teleoperator attends and resumes the autonomous operation.  

\section{VALIDATION AND DISCUSSION}
\label{sec:results}
The feasibility of the proposed digital factory architecture was evaluated in three manufacturing operations: CNC machining, robotic-assisted abrasive finishing, and robotized disassembly. The following three previously
published cyber-physical manufacturing systems each realize
a subset of the proposed architecture. They were developed
and evaluated independently, and were not operated together
as an integrated factory: 
\begin{itemize}
    \item CNC machining~\cite{Kaarlela2025cyber}. 
    \item Robotic-assisten abrasive finishing~\cite{outeiro_development_2025}. 
    \item Robotized EVB disassembly \cite{Kaarlela2025evb}.
\end{itemize}

\begin{figure}[h!tb]
    \centering
    \begin{subfigure}[t]{0.48\columnwidth}
        \centering
        \includegraphics[width=\linewidth]{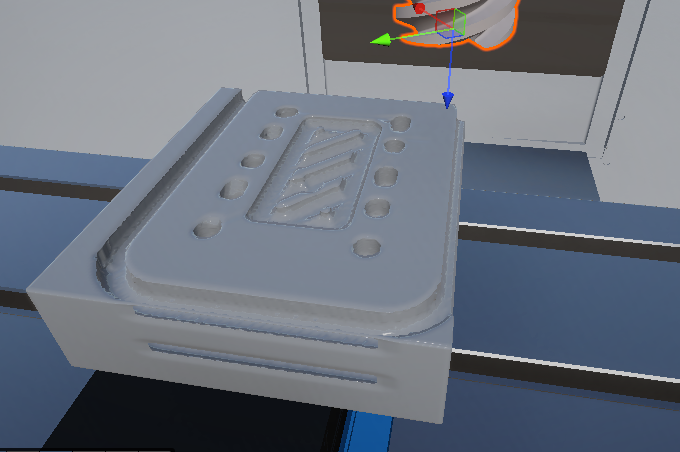}
        \caption{CNC machining process digital twin \cite{Kaarlela2025cyber}.}
        \label{fig:firstW}
    \end{subfigure}
    \hfill
    \begin{subfigure}[t]{0.48\columnwidth}
        \centering
        \includegraphics[width=\linewidth]{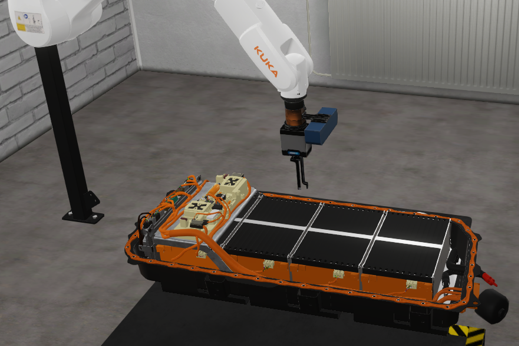}
        \caption{Robotized EVB disassembly \cite{Kaarlela2025evb}.}
        \label{fig:secondW}
    \end{subfigure}
    \vspace{0.5em}
    \begin{subfigure}[t]{0.48\columnwidth}
        \centering
        \includegraphics[width=\linewidth]{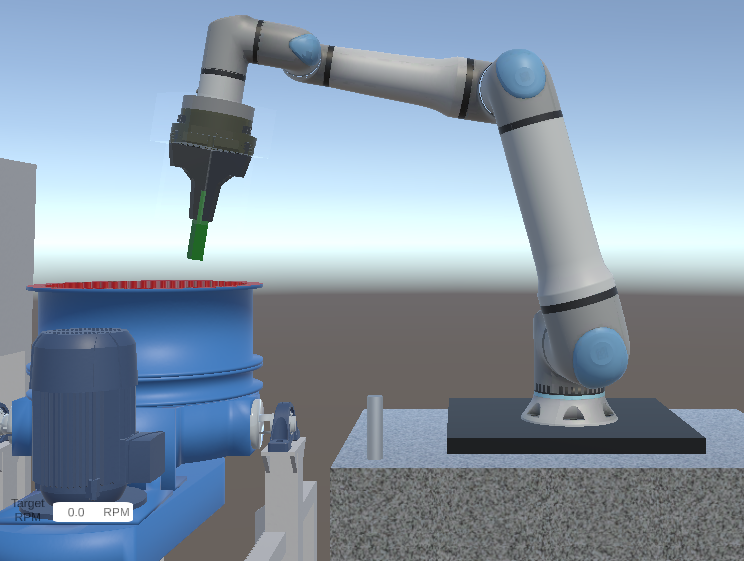}
        \caption{Robotic-Assisted abrasive finishing process \cite{outeiro_development_2025}.}
        \label{fig:thirdW}
    \end{subfigure}

        \caption{Representative manufacturing systems used to validate the proposed digital factory architecture.}
    \label{fig:usecases}
    
\end{figure}

Together, these implementations demonstrate the technological building blocks required for disembodied work. Figure~\ref{fig:usecases} illustrates the corresponding XR teleoperation interfaces of the three manufacturing operations. The implementations originally developed as individual systems create the building blocks for the unified digital factory presented in this paper.

Our study aims to demonstrate the principal capabilities of the proposed digital factory: bidirectional synchronization between physical and digital twins, remote teleoperation by an operator, autonomous operation, cloud-based services, and edge-based AI. Together, these systems provide proof-of-concept evidence that heterogeneous manufacturing resources can be incorporated into a common operational digital factory architecture.

\subsection{Digital Twin Synchronization}
\label{subsec:dtsynchro}
The proposed architecture relies on continuously synchronized DTs to maintain consistency between physical and virtual manufacturing systems. The measured round-trip communication latency for these MQTT-based teleoperation implementations has ranged from 431 ms to 1.6 s, averaging 563 ms, demonstrating the feasibility of synchronized remote operation over cloud-based infrastructure. Together, these implementations demonstrate that manufacturing resources can maintain continuously synchronized DTs within a common digital factory architecture.

For the CNC machining application, a TCP-to-MQTT bridge enabled real-time synchronization of machine coordinates, tool states, spindle speed, and process states with a web-based DT. Similarly, the EVB disassembly system synchronized robot joint states, battery pose, and process execution between physical and DTs before physical execution. For abrasive finishing, the robot and abrasive finishing equipment operational states were synchronized between the twins. These implementations demonstrate that heterogeneous manufacturing systems can maintain continuously synchronized DTs within a digital factory. 

\subsection{Human-AI Collaboration}
\label{subsec:humanaicollaboration}
Humans remain an essential part of the proposed digital factory architecture. The EVB disassembly system was extended~\cite{Karami2025autonomous} to demonstrate a hybrid operating paradigm where human operators initially perform unknown disassembly operations through teleoperation while the resulting task sequences are stored for subsequent autonomous execution.

\begin{figure}[h!tb]
  \centering
 \includegraphics[width=\columnwidth]{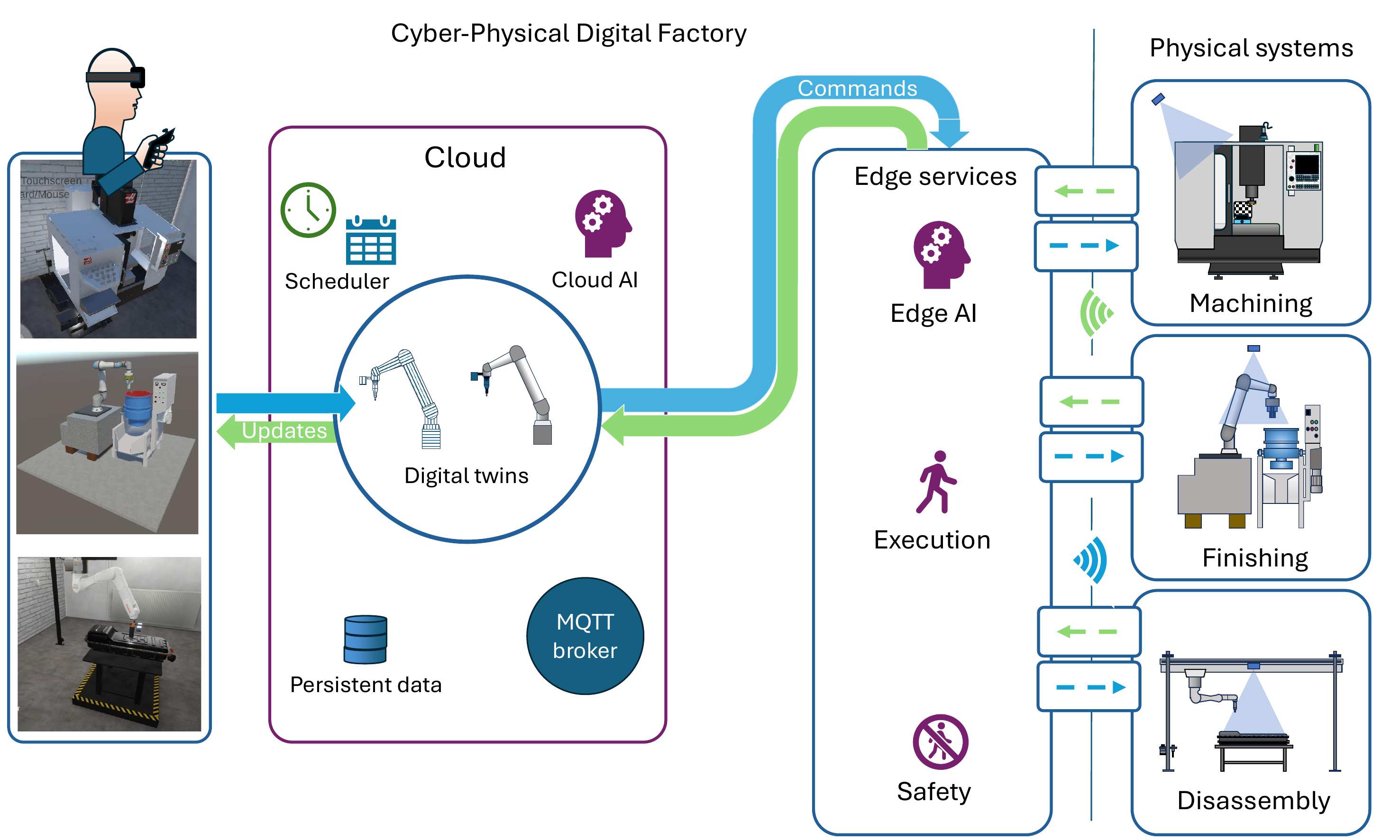}
  \caption{Operational view of the proposed Cyber-Physical Digital Factory. Cloud services coordinate DTs, AI, scheduling, and persistent data; edge services execute manufacturing operations and enforce local safety.}
  \label{fig:architecture2}
\end{figure}

Edge AI further extends autonomous operation by performing machine vision, component recognition, and connector localization directly on the manufacturing system. Connector detection achieved a precision of 98.1\% and recall of 96.5\%, while wire segmentation achieved a precision of 74\% and recall of 82\%. When confidence falls below predefined thresholds, control is transferred seamlessly to the remote operator, illustrating the collaborative human–AI paradigm envisioned by the proposed digital factory. Figure~\ref{fig:architecture2} summarizes the cyber-physical interactions between factory-level services, DTs, autonomous manufacturing systems, and humans. The architecture establishes multiple interacting feedback loops, enabling coordinated decision-making across physical and cyber domains.

\subsection{Validation of the Digital Factory Architecture}
\label{subsec:validationofdtarch}
The previous implementations collectively validate the principal architectural capabilities of the proposed digital factory. DTs provide synchronized representations of manufacturing systems, MQTT enables interoperable communication, cloud services provide remote access and factory-level coordination, edge AI enables autonomous operation close to physical equipment, and teleoperation supports human intervention whenever autonomous execution reaches its operational limits. Table~\ref{tab:validation} summarizes how the implemented systems validate the principal functions of the proposed digital factory. 

\begin{table}[h!tb] 
\caption{Digital factory CPS capabilities validation.} \label{tab:validation} 
\centering 
\begin{tabular}{p{2.9cm} p{5.2cm}} \hline 
\textbf{CPS Property} & \textbf{Validation Evidence} \\ \hline 
State synchronization & MQTT-based DT synchronization with 431 ms--1.6 s round-trip latency (avg. 563 ms). \\ 
Observability & Continuous monitoring of CNC machine and robot states, and process execution using synchronized DTs. \\ 
Controllability & XR-based teleoperation demonstrated for CNC machining and EV battery disassembly. \\
Autonomy & EVB Connector detection precision 98.1\%, recall 96.5\%; wire segmentation precision 74\%, recall 82\%. \\ 
Human-in-the-loop operation & Control transferred from AI to human when confidence thresholds are violated. \\ 
Interoperability & Common MQTT infrastructure integrated CNC machining, robotic finishing, and EVB disassembly. \\ \hline 
\end{tabular} 
\end{table}

Collectively, the presented implementations demonstrate that the proposed digital factory architecture is \textit{application-independent}. Although the manufacturing systems differ substantially in terms of physical processes, sensing modalities, and levels of autonomy, all employ the same architectural principles comprising synchronized DTs, publish/subscribe communication, hierarchical cloud-edge AI, and human supervision and teleoperation. These common characteristics indicate that the proposed architecture can provide a reusable foundation for heterogeneous manufacturing systems supporting disembodied work.

\subsection{Cybersecurity}
\label{subsec:cybersecurity}
Cybersecurity is considered a continuous lifecycle activity rather than a one-time system configuration. Intrusion Detection Systems (IDSs) and Intrusion Prevention Systems (IPSs) can strengthen the platform by providing real-time detection and mitigation of malicious activities, complementing periodic vulnerability scans and software maintenance~\cite{Culot2019addressing, Mullet2021review}. Cyberattacks on connected manufacturing systems may directly affect physical equipment and operator safety; cybersecurity should be regarded as an integral component of safe cyber-physical operation.

\subsection{Current limitations and Future Work}
\label{subsec:futurework}
The presented results should be interpreted as preliminary validation of the architectural components --- rather than as a demonstration of fully integrated factory-wide orchestration. The individual systems were developed and evaluated separately, and continuous coordination between all manufacturing resources through a single factory-level scheduler has not yet been experimentally validated. Thus, the main result of the present work is the synthesis of these validated components into a unified digital factory architecture and the demonstration of their compatibility with the operating principles required for disembodied work.

Implementing a persistent work storage would enable storing information about the manufactured products. The repository stores synchronized operational data, workpiece geometry, manufacturing parameters, anomalies detected, and timestamps. The stored information would represent the complete manufacturing history of an individual workpiece and is associated with its corresponding part identifier. The scalability of the proposed architecture under concurrent operation of multiple manufacturing resources has not yet been experimentally evaluated. Likewise, robustness against communication latency, network disruptions, and dynamic resource allocation requires further investigation in industrial-scale deployments. 

\section{CONCLUSIONS}
This paper proposed a unified digital factory architecture that enables disembodied work by integrating synchronized DTs, hierarchical cloud-edge AI, teleoperation, and autonomous manufacturing resources into a continuously synchronized cyber-physical manufacturing environment. Unlike conventional digital factory concepts primarily supporting production planning and simulation, the proposed architecture provides an operational framework for location-independent manufacturing operations.

The feasibility of the proposed architecture was demonstrated through representative manufacturing applications including CNC machining, robotic-assisted abrasive finishing, and EVB disassembly. Although developed independently, these implementations collectively validated the digital factory key architectural capabilities, including DT synchronization, remote supervision, AI-assisted autonomous operation, and secure cloud-edge communication.

Future work will focus on implementing persistent data storage, integrating factory-wide production scheduling and resource orchestration, and validating the architecture in a multi-cell manufacturing environment that supports real-time coordination across heterogeneous manufacturing systems.






\section*{ACKNOWLEDGMENT}

This research was supported by RIS4E - Revolutionary and 
Intelligent Steel Solutions for Sustainable Environment 
(Business Finland 43/31/2026). Financial support provided by the University of North Carolina at Charlotte through the Center for Precision Metrology Affiliates Program is also acknowledged.

\bibliographystyle{IEEEtran}
\bibliography{citations}

@article{Kaarlela2025cyber,
title = {A cyber-physical machine tool concept for education and workforce training in {CNC} machining},
journal = {Manufacturing Letters},
volume = {44},
pages = {1209-1218},
year = {2025},
note = {53rd SME North American Manufacturing Research Conference (NAMRC 53)},
issn = {2213-8463},
doi = {10.1016/j.mfglet.2025.06.140},
author = {Kaarlela, Tero and Outeiro, José}
}

@article{Debevec2022digital,
title = {Digital factory to support deadlines prediction in small volume production},
journal = {IFAC-PapersOnLine},
volume = {55},
number = {10},
pages = {2306-2311},
year = {2022},
note = {10th IFAC Conference on Manufacturing Modelling, Management and Control MIM 2022},
issn = {2405-8963},
doi = {10.1016/j.ifacol.2022.10.052},
author = {Debevec, M. and Simic, M. and Herakovic, N.}
}

@Article{Denisa2023technology,
AUTHOR = {Deniša, Miha and Ude, Aleš and Simonič, Mihael and Kaarlela, Tero and Pitkäaho, Tomi and Pieskä, Sakari and Arents, Janis and Judvaitis, Janis and Ozols, Kaspars and Raj, Levente and Czmerk, András and Dianatfar, Morteza and Latokartano, Jyrki and Schmidt, Patrick Alexander and Mauersberger, Anton and Singer, Adrian and Arnarson, Halldor and Shu, Beibei and Dimosthenopoulos, Dimosthenis and Karagiannis, Panagiotis and Ahonen, Teemu-Pekka and Valjus, Veikko and Lanz, Minna},
TITLE = {Technology {M}odules {P}roviding {S}olutions for {A}gile {M}anufacturing},
JOURNAL = {Machines},
VOLUME = {11},
YEAR = {2023},
NUMBER = {9},
ARTICLE-NUMBER = {877},
ISSN = {2075-1702}
}

@misc{robert2020growing,
author={Scott, Robert E. and Mokhiber, Zane},
title={Growing {C}hina trade deficit cost 3.7 million {A}merican jobs between 2001 and 2018},
year={2020},
urldate = {2020-01-30},
note = {Accessed: 29th July 2026},
howpublished = {\url{https://www.epi.org/publication/growing-china-trade-deficits-costs-us-jobs/}}
}

@misc{Mohan2024genz,
author = {Mohan, Pavithra},
title  = {Gen {Z} is interested in blue-collar work—but not necessarily manufacturing},
journal = {Business Insider},
year    = {2024},
note = {Accessed: 29th July 2026},
howpublished = {\url{https://www.fastcompany.com/91319434/gen-z-is-interested-in-blue-collar-work-but-not-manufacturing-jobs}}
}

@misc{Stephenson2025digital,
author={Stephenson, David},
title={Digital {N}atives and the {F}uture {W}orkplace},
year={2024},
note = {Accessed: 29th July 2026},
howpublished = {\url{https://www.littleonline.com/insights/digital-natives-and-the-future-workplace/}}
}

@misc{Paradis2024genz,
author = {Paradis, Tim},
title  = {Gen {Z} isn't lazy — it just needs training, {HR} experts say},
journal = {Business Insider},
year    = {2024},
note = {Accessed: 29th July 2026},
howpublished = {\url{https://www.businessinsider.com/gen-z-lazy-hiring-jobs-careers-hr-training-pandemic-2024-8}}
}

@misc{bartik2023rise,
author={Bartik, Alexander and Cullen, Zoe and Glaeser, Ed and Luca, Michael and Stanton, Christopher},
title={The {R}ise of {R}emote {W}ork: {E}vidence on {P}roductivity and {P}references from {F}irm and {W}orker {S}urveys},
year={2023},
urldate = {2020-01-30},
note = {Accessed: 29th July 2026},
howpublished = {\url{https://www.hbs.edu/faculty/Pages/download.aspx?name=20-138.pdf}}
}

@article{Bracht2005the,
title = {The {D}igital {F}actory between vision and reality},
journal = {Computers in Industry},
volume = {56},
number = {4},
pages = {325-333},
year = {2005},
note = {The {D}igital {F}actory: {A}n {I}nstrument of the {P}resent and the {F}uture},
issn = {0166-3615},
doi = {10.1016/j.compind.2005.01.008},
author = {Bracht, U. and Masurat, T.}
}

@Inbook{Bracht2018digitale,
author={Bracht, Uwe and Geckler, Dieter and Wenzel, Sigrid},
title={Einleitung, {D}efinition und {S}tand der {U}msetzung sowie der {B}ezug zu {I}ndustrie 4.0},
bookTitle={Digitale Fabrik: Methoden und Praxisbeispiele},
year={2018},
publisher={Springer Berlin Heidelberg},
address={Berlin, Heidelberg},
pages={1--25},
isbn={978-3-662-55783-9},
doi={10.1007/978-3-662-55783-9_1}
}

@InProceedings{Karami2025autonomous,
author={Karami, Narjes and Pitk{\"a}aho, Tomi and Kaarlela, Tero},
editor={Huber, Marco and Verl, Alexander and Kraus, Werner},
title={Autonomous {R}obotized {D}etachment of {W}iring {C}onnectors},
booktitle={European Robotics Forum 2025},
year={2025},
publisher={Springer Nature Switzerland},
address={Cham},
pages={288--293},
isbn={978-3-031-89471-8}
}

@misc{wullf2024rise,
      title={The rise in remote work since the pandemic and its impact on productivity}, 
      author={Sabrina Wulff Pabilonia and Jill Janocha Redmond},
      year={2024},
      howpublished ={\url{https://www.bls.gov/opub/btn/volume-13/remote-work-productivity.htm}}, 
}

@misc{Moffat2021big,
author = {Moffat, Mike},
title  = {The big shift: {C}hanges in Canadian manufacturing employment, 2003-2018},
journal = {Future Skills Centre},
year    = {2021},
note = {Accessed 28 July 2026},
howpublished = {\url{https://fsc-ccf.ca/research/the-big-shift-changes-in-canadian-manufacturing-employment-2003-2018-full-report}}
}

@ARTICLE{Jin2023achieving,
  author={Jin, Ziyue and Marian, Romeo M. and Chahl, Javaan S.},
  journal={The International Journal of Advanced Manufacturing Technology}, 
  title={Achieving batch-size-of-one production model in robot flexible assembly cells}, 
  year={2023},
  volume={5},
  number={},
  pages={2097-2116},
  doi = {10.1007/s00170-023-11246-y}
}

@article{Burggraf2024towards,
title = {Towards {D}igital-{T}win-Driven {F}actory {P}lanning – {A} {S}ystematic {R}eview},
journal = {Procedia CIRP},
volume = {126},
pages = {248-253},
year = {2024},
note = {17th CIRP Conference on Intelligent Computation in Manufacturing Engineering (CIRP ICME ‘23)},
issn = {2212-8271},
doi = {10.1016/j.procir.2024.08.334},
author = {Burggräf, Peter and Adlon, Tobias and Schäfer, Niklas}
}

@article{Tao2024advancements,
author = {Tao, Fei and Zhang, He and Zhang, Chenyuan}, 
year ={2024},
title = {Advancements and challenges of digital twins in industry},
journal = {Nature Computational Science},
pages ={169-177},
volume ={4},
issue ={3},
issn ={2662-8457},
doi ={10.1038/s43588-024-00603-w}
}

@article{Eichelberger2025industry,
title = {Industry 4.0/{II}o{T} {P}latforms for manufacturing systems — {A} systematic review contrasting the scientific and the industrial side},
journal = {Information and Software Technology},
volume = {179},
pages = {107650},
year = {2025},
issn = {0950-5849},
doi = {10.1016/j.infsof.2024.107650},
author = {Eichelberger, Holger and Sauer, Christian and Ahmadian, Amir Shayan and Kröher, Christian}
}

@article{Kim2024teleoperator,
  author = {Kim, Sunwook and Hernandez, Ivan and Nussbaum, Maury A. and Lim, Sol},
  title = {Teleoperator-{R}obot-{H}uman {I}nteraction in {M}anufacturing: {P}erspectives from {I}ndustry, {R}obot {M}anufacturers, and {R}esearchers},
  journal = {IISE Transactions on Occupational Ergonomics and Human Factors},
  volume = {12},
  number = {1-2},
  pages = {28--40},
  year = {2024},
  doi = {10.1080/24725838.2024.2310301}
}

@article{Wan2024virtual,
title = {A virtual reality-based immersive teleoperation system for remote human-robot collaborative manufacturing},
journal = {Manufacturing Letters},
volume = {41},
pages = {43-50},
year = {2024},
note = {52nd SME North American Manufacturing Research Conference (NAMRC 52)},
issn = {2213-8463},
doi = {10.1016/j.mfglet.2024.09.008},
author = {Wan, Ke and Li, Chengxi and Lo, Fo-Sing and Zheng, Pai}
}

@INPROCEEDINGS{Kaarlela2025evb,
  author={Kaarlela, Tero and Salo, Sami and Outeiro, Jose},
  booktitle={2025 IEEE 21st International Conference on Automation Science and Engineering (CASE)}, 
  title={Digital twin and extended reality for teleoperation of the electric vehicle battery disassembly}, 
  year={2025},
  volume={},
  number={},
  pages={3393-3398},
  doi={10.1109/CASE58245.2025.11163970}
  }

@inproceedings{outeiro_development_2025,
	address = {Porto, Portugal},
	title = {Development of a {D}igital {T}win of an {I}ntelligent {R}obot-{A}ssisted {F}inishing {S}ystem for {P}olishing {M}etal {A}dditive {M}anufactured {C}omponents},
	booktitle = {Proceedings of the 8th {I}nternational {C}onference on {I}ntegrity, {R}eliability and {F}ailure of {E}ngineering {S}ystems and {M}aterials ({IRF2025})},
	author = {Outeiro, Jose and Holt, Jia and Kaarlela, Tero and Uçak, Necati and Greis, Noel and Cherukuri, Harish},
	month = {jul},
	year = {2025}
}

@ARTICLE{Mullet2021review,
    author={Mullet, Valentin and Sondi, Patrick and Ramat, Eric},
    journal={IEEE Access}, 
    title={A {R}eview of {C}ybersecurity {G}uidelines for {M}anufacturing {F}actories in {I}ndustry 4.0}, 
    year={2021},
    volume={9},
    number={},
    pages={23235-23263},
    doi={10.1109/ACCESS.2021.3056650}
    }

@ARTICLE{Culot2019addressing,  
author={Culot, Giovanna and Fattori, Fabio and Podrecca, Matteo and Sartor, Marco},
journal={IEEE Engineering Management Review},
title={Addressing {I}ndustry 4.0 {C}ybersecurity {C}hallenges},
year={2019},
volume={47},
number={3},
pages={79-86},
doi={10.1109/EMR.2019.2927559}
}

@misc{ISO23247part2,
  author = {{International Organization for Standardization}},
  title  = {{ISO} 23247-2:2021 Automation systems and integration --- Digital twin framework for manufacturing --- Part 2: Reference architecture},
  year   = {2021},
  howpublished = {Standard, ISO, Geneva, Switzerland}
}

@article{Lee2015cyber,
  author  = {Lee, Jay and Bagheri, Behrad and Kao, Hung-An},
  title   = {A Cyber-Physical Systems architecture for {I}ndustry 4.0-based manufacturing systems},
  journal = {Manufacturing Letters},
  volume  = {3},
  pages   = {18--23},
  year    = {2015},
  doi     = {10.1016/j.mfglet.2014.12.001}
}

@misc{IEC63278,
  author = {{International Electrotechnical Commission}},
  title  = {{IEC} 63278-1 Asset {A}dministration {S}hell for industrial applications --- Part 1: Asset Administration Shell structure},
  year   = {2023},
  howpublished = {Standard, IEC, Geneva, Switzerland}
}

@misc{IEC62264,
  author = {{International Electrotechnical Commission}},
  title  = {{IEC} 62264-1 Enterprise-control system integration --- Part 1: Models and terminology},
  year   = {2013},
  howpublished = {Standard, IEC, Geneva, Switzerland}
}

@INPROCEEDINGS{simplex_lui_sha,
  author={Seto, D. and Krogh, B. and Sha, L. and Chutinan, A.},
  booktitle={Proceedings of the 1998 American Control Conference. ACC (IEEE Cat. No.98CH36207)}, 
  title={The Simplex architecture for safe online control system upgrades}, 
  year={1998},
  volume={6},
  number={},
  pages={3504-3508 vol.6},
  doi={10.1109/ACC.1998.703255}}

@inproceedings{sridhar2025regent,
  title     = {{REGENT}: A Retrieval-Augmented Generalist Agent That Can Act In-Context in New Environments},
  author    = {Sridhar, Kaustubh and Dutta, Souradeep and Jayaraman, Dinesh and Lee, Insup},
  booktitle = {International Conference on Learning Representations (ICLR)},
  year      = {2025},
  pages     = {39385--39414},
}

\end{document}